\documentstyle[tighten,preprint,eqsecnum,%
aps,prd]{revtex}
\begin{document}
\draft

\hyphenation{
mani-fold
mani-folds
}




\def\casehalf{{\case{1}{2}}}

\def\ads{{anti-de~Sitter}}
\def\rnads{{Reissner-Nordstr\"om-anti-de~Sitter}}
\def\rn{{Reissner-Nordstr\"om}}

\def\bm{{\bf m}}
\def\bq{{\bf q}}
\def\br{{\bf r}}

\def\Rh{R_{\rm h}}
\def\Rhor{R_{\rm hor}}
\def\Rcrit{R_{\rm crit}}


\preprint{\vbox{\baselineskip=12pt
\rightline{Freiburg THEP-98/18}
\rightline{gr-qc/9809005}}}
\title{Hamiltonian evolution and quantization 
for extremal black holes}
\author{Claus Kiefer\footnote{%
Electronic address:
Claus.Kiefer@physik.uni-freiburg.de
}}
\address{
Fakult\"at f\"ur Physik, 
Universit\"at Freiburg, 
Hermann-Herder-Stra\ss{}e~3, 
D--79104 Freiburg, 
Germany}
\author{Jorma Louko\footnote{%
Electronic address:
louko@aei-potsdam.mpg.de
}}
\address{
Max-Planck-Institut f\"ur Gravitations\-physik,
Schlaatzweg~1,
D--14473 Potsdam,
Germany
}
\date{Revised version, September 1998}
\maketitle
\begin{abstract}%
  We present and contrast two distinct ways of including extremal
  black holes in a Lorentzian Hamiltonian quantization of spherically
  symmetric Einstein-Maxwell theory. First, we formulate the classical
  Hamiltonian dynamics with boundary conditions appropriate for 
  extremal black holes only. The Hamiltonian contains no surface term
  at the internal infinity, for reasons related to the vanishing of
  the extremal hole surface gravity, and quantization yields a
  vanishing black hole entropy. Second, we give a Hamiltonian
  quantization that incorporates extremal black holes as a limiting
  case of nonextremal ones, and examine the classical limit in terms
  of wave packets. The spreading of the packets, even the ones
  centered about extremal black holes, is consistent with continuity
  of the entropy in the extremal limit, and thus with the
  Bekenstein-Hawking entropy even for the extremal holes. The
  discussion takes place throughout within Lorentz-signature
  spacetimes.
\end{abstract}

\narrowtext

\section{Introduction}
\label{sec:intro}

Extremal black holes have a special and controversial 
status in black hole
thermodynamics. On the one hand, one might expect macroscopic
thermodynamical quantities to be continuous functions of the black
hole parameters when one passes from nonextremal black holes to the
extremal limit, and one would then be led to the conclusion that
extremal black holes must posses the standard Bekenstein-Hawking
entropy, equal to one quarter of the horizon area
\cite{bekenstein1,bekenstein2,hawkingCMP,davies1,davies2}. Support for
this view is also provided by state-counting calculations of certain
extremal and near-extremal black holes in string theory; for a review,
see Ref.\ \cite{Peet}.  On the other hand, extremal and nonextremal
black hole geometries have qualitative differences that raise 
doubts about limiting arguments. In particular, the 
absence of a bifurcate Killing horizon and the vanishing of the
horizon surface gravity in
the Lorentzian extremal black hole 
spacetime furnish the Riemannian section of
the spacetime with
certain properties that, within the path-integral approach to
black-hole thermodynamics \cite{GH1,hawkingCC,pagerev}, 
have lead to 
the conclusion of a vanishing
entropy \cite{haw-horo-ross,teitel-entropy}.  

In this paper we make two observations that aim to clarify the
distinctive status of extremal black hole geometries within
Hamiltonian quantization.  We consider the limited but
concrete context of spherically symmetric geometries in four spacetime
dimensions, within the Einstein-Maxwell theory.  

First, we formulate the classical 
Hamiltonian dynamics of Lorentzian spacetimes under boundary
conditions appropriate for {\em only\/} extremal black holes, such
that the spacelike hypersurfaces are asymptotic to hypersurfaces of
constant Killing time at the internal infinity. We show that the
action does not depend on how, or whether, the evolution of the
spacelike hypersurfaces near the internal infinity is prescribed in
the variational principle, and we trace the geometrical reason for
this directly to the vanishing of the surface gravity. We also find
explicitly the reduced Hamiltonian theory, with a two-dimensional
reduced phase space. The action does not contain at the internal
infinity a boundary term of the kind that gives rise to the
Bekenstein-Hawking entropy in the Hamiltonian quantization of
nonextremal black holes 
\cite{LW2,lau1,BLPP,lou-win,love-hole,brotz-kiefer,Br,Ki}. 
Quantizing this Hamiltonian theory on its own would therefore 
lead to the conclusion of 
a vanishing extremal black hole entropy. 

Second, we formulate a Hamiltonian quantization that incorporates 
extremal black holes as a limiting case of nonextremal ones, 
and we explore
in this quantization wave packets centered around a
given black hole. The wave packets centered around extremal and
non-extremal holes have some qualitative differences, in particular in
how they spread as a function of time. However, the spreading behavior
also suggests that in this approach both extremal and
nonextremal holes should have the standard Bekenstein-Hawking entropy.
Physically, for a wave packet centered about an extremal hole,
the extremal configuration itself is a set of measure zero, and the
entropy should then in essence be determined by the nonextremal
configurations in a small but finite neighborhood of the packet
center.

Neither of these results is surprising, and they are fully in
line with the general expectation that treating extremal black
holes on their own lead to a vanishing entropy while treating them as
limiting cases of nonextremal ones lead to the usual
Bekenstein-Hawking entropy.
However, with their express emphasis 
on a Lorentz-signature spacetime and on a Hamiltonian
framework of quantization, we view our results as 
filling a gap in the literature. 
Concurring observations were recently
made, within a technically different approach that relies on 
Euclidean-signature spacetimes, in
Ref.\ \cite{mitra}. 

The rest of the paper is as follows. In section
\ref{sec:fall-ex-hor} we set up
a classical Hamiltonian theory under boundary conditions appropriate
for extremal black holes, and in section \ref{sec:extr-reduction} 
we reduce this theory to
its unconstrained, two-dimensional phase space. 
A~Hamiltonian
quantization including both extremal and nonextremal holes is analyzed
in section~\ref{sec:packets}. 
Section \ref{sec:conclusion} presents brief concluding remarks. Some
relevant properties of the \rnads\ family of spacetimes are collected
in the Appendix. We use the geometrical units \cite{MTW} in which 
$c = G = 1$, but we keep Planck's constant~$\hbar$.

\section{Canonical formalism for extremal holes}
\label{sec:fall-ex-hor}

In this section we present a Hamiltonian formulation for spherically
symmetric Einstein-Maxwell spacetimes under boundary conditions
appropriate for the exterior region of an extremal black hole, such
that the spacelike hypersurfaces are asymptotic to the hypersurfaces
of constant Killing time near the internal infinity. For concreteness,
we consider a negative cosmological constant; the cases of a
positive or vanishing cosmological constant can be handled with
straightforward modifications and similar conclusions. As our new results
concern just the internal infinity, we
leave the boundary conditions and boundary terms 
at the ``outer'' end of the spacelike 
hypersurfaces unspecified. Examples of how to 
handle the outer end can be found in 
Refs.\ \cite{LW2,lau1,BLPP,lou-win,love-hole,brotz-kiefer,Br,Ki}. 

The notation follows Ref.\ \cite{lou-win}.  
We write the cosmological
constant as~$-3 \ell^{-2}$, where $\ell>0$.  

We consider the general spherically symmetric ADM metric
\begin{equation}
ds^2 = - N^2 dt^2 + \Lambda^2 {(dr + N^r dt)}^2 +R^2 d\Omega^2
\ \ ,
\label{4-metric}
\end{equation}
where $d\Omega^2$ is the metric on the unit two-sphere, and $N$,
$N^r$, $\Lambda$ and $R$ are functions of $t$ and~$r$. We take
the electromagnetic bundle to be trivial, and describe the
electromagnetic field by the globally-defined spherically symmetric
one-form
\begin{equation}
A = \Gamma dr + \Phi dt
\ \ ,
\label{4-vectorpot}
\end{equation}
where $\Gamma$ and $\Phi$ are functions of $t$ and~$r$.  The
coordinate $r$ takes the semi-infinite range $[0,\infty)$. We assume
both the spatial metric and the spacetime metric to be nondegenerate,
and $\Lambda$, $R$, and $N$ to be positive. 

The bulk contribution to the Hamiltonian action of Einstein-Maxwell
theory reads
\begin{equation}
S_\Sigma 
= \int dt
\int_0^\infty dr \left( P_\Lambda {\dot \Lambda} +
P_R {\dot R} + P_\Gamma {\dot \Gamma}
- NH - N^r H_r - {\tilde \Phi} G
\right)
\ \ ,
\label{S-ham}
\end{equation}
where the super-Hamiltonian constraint~$H$, the radial supermomentum
constraint~$H_r$, and the Gauss law constraint $G$ are given by
\begin{mathletters}
\label{constraints}
\begin{eqnarray}
H &=& - R^{-1} P_R P_\Lambda
+ \case{1}{2} R^{-2} \Lambda
( P_\Lambda^2 + P_\Gamma^2)
\nonumber
\\
&&
+ \Lambda^{-1} R R'' - \Lambda^{-2} R R' \Lambda'
+ \case{1}{2} \Lambda ^{-1} {R'}^2
- \case{1}{2} \Lambda
- \case{3}{2} \ell^{-2} \Lambda R^2
\ \ ,
\label{superham}
\\
H_r &=& P_R R' - \Lambda P_\Lambda' - \Gamma P_\Gamma'
\ \ ,
\label{supermom}
\\
G &=& - P_\Gamma'
\ \ ,
\label{gauss}
\end{eqnarray}
\end{mathletters}%
and we have defined 
\begin{equation}
{\tilde \Phi} := \Phi - N^r \Gamma
\ \ . 
\label{Phi-redef}
\end{equation}
We regard $N$, $N^r$, and ${\tilde \Phi}$ as the independent Lagrange
multipliers in the action~(\ref{S-ham}). Local variation of
(\ref{S-ham}) yields the three constraint equations $H=H_r=G=0$, 
and the six dynamical equations that give the time derivatives of the
coordinates and the momenta \cite{lou-win}. These equations correctly
reproduce the spherically symmetric Einstein-Maxwell equations.

We now wish to adopt near $r=0$ boundary conditions that enforce every
classical solution to be (part of) an exterior region of an extremal
\rnads\ (RNAdS) black hole (see the Appendix), 
and such that the constant $t$
hypersurfaces are asymptotic to the constant Killing time
hypersurfaces at the internal infinity as $r\to0$. 
To this end, we take the variables to have the small $r$ expansion 
\begin{mathletters}
\label{s-r}
\begin{eqnarray}
\Lambda (t,r) &=& \Lambda_{-1}(t) r^{-1} + O(1)
\ \ ,
\label{s-r-Lambda}
\\
R(t,r) &=& R_0(t) + R_1(t)r +  O(r^2)
\ \ ,
\label{s-r-R}
\\
P_{\Lambda}(t,r) &=& O(r^3)
\ \ ,
\label{s-r-PLambda}
\\
P_{R}(t,r) &=& O(r)
\ \ ,
\label{s-r-PR}
\\
N(t,r) &=& 
\Lambda^{-1} R' \left( {\tilde N}_0(t) + O(r) \right) 
\ \ ,
\label{s-r-N}
\\
N^r(t,r) &=& O(r)
\ \ ,
\label{s-r-Nr}
\\
\Gamma(t,r) &=& \Gamma_0(t) + O(r)
\ \ ,
\label{s-r-B}
\\
P_\Gamma(t,r) &=& Q_0(t) + O(r^2) 
\ \ ,
\label{s-r-PB}
\\
{\tilde \Phi}(t,r) &=& {\tilde \Phi}_0(t) + O(r)
\ \ ,
\label{s-r-tPhi}
\end{eqnarray}
\end{mathletters}
where $\Lambda_{-1}$, $R_0$, and $R_1$ are positive. Here $O(r^n)$
stands for a term whose magnitude at $r\to0$ is bounded by $r^n$ times
a constant, and whose derivatives with respect to $r$ and $t$ fall off
accordingly. It is straightforward to verify that the falloff
conditions (\ref{s-r}) are consistent with the
constraints, and that they are preserved by the
time evolution equations when the constraints hold for the initial
data.

To see that the falloff (\ref{s-r}) accomplishes what we wish, 
consider a classical solution satisfying this
falloff. First, as $R_1$ is positive, the radius of the
two-sphere is not constant on the spacetime, and the 
solution cannot belong to the Bertotti-Robinson-type 
family \cite{exact-book}. The solution is therefore part of a 
RNAdS spacetime. 
Second, recall that on a classical solution, the
functions $F$ and $T$ defined in the 
Appendix can be written in terms of the canonical variables as
\cite{kuchar1,lou-win}
\begin{mathletters}
\label{F-and-Tprime}
\begin{eqnarray}
F &=& {\left( {R' \over \Lambda} \right)}^2
- {\left( { P_\Lambda \over R} \right)}^2
\ \ ,
\label{F-def-met}
\\
-T' &=& R^{-1} F^{-1} \Lambda P_\Lambda
\ \ , 
\label{Tprime}
\end{eqnarray}
\end{mathletters}
for which equations (\ref{s-r}) imply the falloff 
\begin{mathletters}
\begin{eqnarray}
F &=& 
\frac{R_1^2 r^2}{\Lambda_{-1}^{2}}  + O(r^3)
\ \ ,
\label{F-fall}
\\
T' &=& O(1)
\ \ .
\label{Tprime-fall}
\end{eqnarray}
\end{mathletters}
Equation (\ref{Tprime-fall}) implies 
that the $r\to0$ end of a
constant $t$ hypersurface tends to a finite value of the Killing time.
Third, equation (\ref{s-r-Lambda}) shows that the proper distance on a
constant $t$ hypersurface from any positive value of $r$ to $r=0$ is
infinite, while equation (\ref{s-r-R}) shows that $R$ must tend to a
finite and nonzero value. 
These properties imply that the solution, if it exists,
is an extremal RNAdS hole, and that 
the constant $t$ hypersurfaces are asymptotic to the constant
Killing time hypersurfaces at the internal infinity as $r\to0$. 
Finally, it is easy to verify that the extremal RNAdS black 
hole {\em can\/} be written in coordinates satisfying~(\ref{s-r}); for
example, a static gauge satisfying (\ref{s-r}) is obtained from the
curvature coordinates of the Appendix via 
$T=t$ and $R = M(1+r)$. 
The falloff (\ref{s-r}) thus has the desired properties. 

Note that when the equations of motion hold, $Q_0$ is the charge
parameter and $R_0$ is the horizon radius, and for an extremal black
hole these are related by equation (\ref{Rcrit}) of the
Appendix: this relation arises in the Hamiltonian formulation as the
leading-order term in the small $r$ expansion of the constraint
equation $H=0$. 
Note also from (\ref{s-r}), (\ref{F-fall}), and (\ref{curv-metric})
that for a classical solution ${\tilde N}_0$ is 
equal to $dT/dt$ at $r=0$. ${\tilde \Phi}_0$~is related to the
electromagnetic gauge at $r=0$ \cite{lou-win}.

Consider now the variational principle. 
The variation of the bulk action
(\ref{S-ham}) contains a volume term
proportional to the equations of motion, 
boundary terms from the
initial and final hypersurfaces, 
and boundary terms from $r=0$ and
$r=\infty$. With the falloff~(\ref{s-r}), 
the boundary term from $r=0$ is 
$- \int dt \, {\tilde \Phi}_0 \, \delta Q_0$. 
Therefore, adding to 
(\ref{S-ham}) the boundary term 
\begin{equation}
\int dt \,
{\tilde \Phi}_0 \, Q_0 
\label{bt-zero}
\end{equation}
(plus appropriate boundary terms at $r=\infty$) yields a
consistent action functional for prescribing ${\tilde \Phi}_0(t)$ 
(and the appropriate quantities at $r=\infty$). 

We emphasize that one does {\em not\/} 
need to add to the action a boundary
term that would refer to the small $r$ behavior of~$N(t,r)$.  The 
action functional is consistent, without any change in the
boundary terms, whether or not one chooses to restrict the variations
of $N(t,r)$ at $r=0$ in some fashion, such as by 
prescribing~${\tilde N}_0(t)$. 
This is a crucial difference between the extremal and nonextremal
black hole variational principles. For a
nonextremal hole, with boundary conditions that make the spacelike
hypersurfaces at $r=0$ asymptotic to constant Killing
time hypersurfaces at the bifurcation two-sphere
\cite{LW2,lau1,BLPP,lou-win,love-hole,brotz-kiefer,Br,Ki}, 
the variation of 
(\ref{S-ham}) contains at $r=0$ also the boundary term 
\begin{equation}
- \casehalf \int dt \, 
\left[ (N/\Lambda)' 
\, \delta (R^2) 
\right]_{r=0}
\ \ . 
\label{nonext-bt-hor}
\end{equation}
Now, both the extremal and nonextremal 
classical black hole solutions satisfy
$[(N/\Lambda)']_{r=0} 
= 
\kappa (dT/dt)_{r=0}$, where 
$\kappa$ is the surface gravity of the hole with respect to the Killing
field~$\partial_T$; one way to see this is to use the fact that, 
in terms of
the function $F(R)$ (\ref{rnads-F}) given in the Appendix, 
$\kappa = \casehalf [\partial F(R)/\partial R]_{R=R_0}$. 
The geometrical reason 
why the surface term (\ref{nonext-bt-hor}) is not present 
under 
the extremal hole falloff is therefore the vanishing
surface gravity of the extremal hole. 

We also emphasize that our falloff (\ref{s-r}) is not a
special case of the nondegenerate horizon falloff adopted
in Refs.\ 
\cite{LW2,lau1,BLPP,lou-win,love-hole,brotz-kiefer,Br,Ki}. 
Rather, the
conditions (\ref{s-r}) imply at the very outset the distinctive
horizon characteristics of the extremal hole, including the vanishing
of the surface gravity. We shall return to this issue in sections
\ref{sec:packets} and~\ref{sec:conclusion}.

\section{Hamiltonian reduction}
\label{sec:extr-reduction}

We wish to eliminate the constraints from the Hamiltonian theory
formulated in section \ref{sec:fall-ex-hor} and express the reduced
theory in terms of an explicit canonical chart. As the dynamical
content of the theory depends on the boundary conditions, we now make
a concrete choice for the falloff at $r\to\infty$, taking the spatial
hypersurfaces there to be asymptotic to hypersurfaces of constant
Killing time as in Ref.\ \cite{lou-win}. We follow closely the method
of Ref.\ \cite{lou-win}, which is an adaptation for the formalism
developed for spherically symmetric vacuum geometries by Kucha\v{r}
\cite{kuchar1}.\footnote{Related reduction methods have been
previously and subsequently considered  in a variety of contexts; 
in addition to Refs.\ 
\cite{LW2,lau1,BLPP,lou-win,love-hole,brotz-kiefer,Br,Ki}, 
see in particular 
Refs.\ 
\cite{bcmn,lund,unruh,fischler,%
thiemann1,thiemann2,thiemann3,thiemann4}. A~more extensive 
list of references is given in Ref.\ \cite{lou-whit-fried}.} 
We shall not aim at a self-contained
presentation, but we shall elaborate on the steps where the new
boundary conditions bring about new features.

The total action consists of the bulk action (\ref{S-ham}) 
and the boundary action given by 
\begin{equation}
S_{\partial\Sigma} :=
\int dt 
\left( 
{\tilde \Phi}_0 Q_0
- {\tilde \Phi}_+ Q_+
- {\tilde N}_+ M_+
\right)
\ \ .
\label{S-boundary}
\end{equation}
The first term in (\ref{S-boundary}) is the boundary
term~(\ref{bt-zero}). In the second term, ${\tilde \Phi}_+$ and $Q_+$
are the asymptotic values of respectively ${\tilde \Phi}$ and
$P_\Gamma$ as $r\to\infty$. The quantity ${\tilde N}_+$ in the third
term characterizes the asymptotic evolution of the spacelike
hypersurfaces at the infinity: on a classical solution, the value of
$dT/dt$ at $r\to\infty$ is equal to~${\tilde N}_+$. Finally, $M_+$~is
the asymptotic value of a phase space function whose value on the
classical solution is just the mass parameter. This action is
appropriate for a variational principle that fixes ${\tilde \Phi}_0$
at $r=0$, and ${\tilde \Phi}_+$ and ${\tilde N}_+$ at $r\to\infty$
\cite{lou-win}. 

As a first step, we make 
a Kucha\v{r}-type \cite{kuchar1} canonical transformation 
from the phase space chart 
$\left\{\Lambda, R, \Gamma, P_\Lambda, P_R, P_\Gamma \right\}$ 
to the new chart
$\left\{M, {\sf R}, Q, P_M,  P_{\sf R}, P_Q\right\}$
by equations (3.7) of Ref.\ \cite{lou-win}. 
It is easy to find the falloff of the new variables, and to verify
that the transformation is canonical under our boundary conditions. 
Of the geometrical meaning of the new variables,
it is here sufficient to recall that on a classical solution, 
$M$ and $Q$ are constants whose values are just the mass and
charge parameters of the spacetime. 

The bulk action in the new chart reads
\begin{equation}
S_\Sigma
= \int dt
\int _0^\infty dr
\left(
P_M { \dot M}
+ P_{\sf R} \dot{\sf R}
+ P_Q {\dot Q}
- N^M M' 
- N^Q Q'
-N^{\sf R} P_{\sf R}
\right)
\ \ ,
\label{Ssigma-new}
\end{equation}
where $N^M$, $N^Q$, and $N^{\sf R}$ are a set of new Lagrange
multipliers, related to the old ones by equations (3.16) of Ref.\
\cite{lou-win}. We are interested in the boundary terms in the 
variation of~(\ref{Ssigma-new}). 
At $r\to\infty$, the situation is as
in Ref.\ \cite{lou-win}: the asymptotic values of
$M$, $Q$, $N^M$, and $N^Q$ are respectively $M_+$, $Q_+$, 
$-{\tilde N}_+$, and~$-{\tilde \Phi}_+$, 
and the boundary term in the
variation of (\ref{Ssigma-new}) at $r\to\infty$ is 
$\int dt \left(
{\tilde N}_+ \delta M_+
+ {\tilde \Phi}_+ \delta Q_+ \right)$. 
At $r\to0$, on the other hand, we have 
\begin{mathletters}
\label{new-small-fall}
\begin{eqnarray}
M  
&=&
M_0 
+
M_1 r 
+ O(r^2)
\ \ ,
\label{new-small-fall-M}
\\
Q 
&=&
Q_0 + O(r^2)
\ \ ,
\\
N^M 
&=&
- {\tilde N}_0 + O(r)
\ \ ,
\\
N^Q 
&=&
- {\tilde \Phi}_0
+ (Q_0/R_0) {\tilde N}_0
+ O(r)
\ \ ,
\end{eqnarray}
\end{mathletters}%
where 
\begin{mathletters}
\label{Mscoeffs}
\begin{eqnarray}
M_0
&=&
\frac{R_0}{2}
\left(
\frac{R_0^2}{\ell^2} + 1 
+ \frac{Q_0^2}{R_0^2}
\right)
\ \ ,
\label{Mscoeff0}
\\
M_1
&=&
\frac{R_1}{2}
\left(
\frac{3 R_0^2}{\ell^2} + 1 
- \frac{Q_0^2}{R_0^2}
\right)
\ \ .
\label{Mscoeff1}
\end{eqnarray}
\end{mathletters} 
The boundary term from $r=0$ in the variation of
(\ref{Ssigma-new}) therefore reads 
\begin{equation}
-
\int dt 
\left[
\frac{{\tilde N}_0}{2}
\left(
\frac{3 R_0^2}{\ell^2} + 1 
- \frac{Q_0^2}{R_0^2}
\right)
\delta R_0
+ {\tilde \Phi}_0 \delta Q_0
\right]
\ \ .
\label{newsboun-first}
\end{equation}
The first term under the integral in (\ref{newsboun-first}) is
proportional to~$M_1$, which vanishes when the bulk constraint
equation $M'=0$ holds.\footnote{Note that $M_1=0$ is equivalent to the
  equation (\ref{Rcrit}) for the horizon radius of an extremal hole.
  The extremality condition thus emerges in the new variables as the
  leading-order term in the small $r$ expansion of the constraint
  $M'=0$, just as it did in the old variables as the leading-order
  term in the small $r$ expansion of the constraint $H=0$.  In the
  next-to-leading order in~$r$, each of these constraint equations can
  be verified to imply the relation $(R_0/\Lambda_{-1})^2 =
  (Q_0/R_0)^2 + 3 (R_0/\ell)^2$.}  This first term in
(\ref{newsboun-first}) therefore vanishes as a consequence of the bulk
variational equations, and only the second term in
(\ref{newsboun-first}) remains.

Collecting these
observations, we see that when ${\tilde N}_+$, ${\tilde \Phi}_+$, and
${\tilde \Phi}_0$ are prescribed, the boundary action to be added to
the bulk action (\ref{Ssigma-new}) is again given
by~(\ref{S-boundary}).  We emphasize that, as in
section~\ref{sec:fall-ex-hor}, this conclusion is 
independent of whether the variation of ${\tilde N}_0$ might
also be restricted in some way.

A Hamiltonian reduction in the new variables is
straightforward. The constraints $M'=0$ and $Q'=0$ imply 
$M(t,r)= \bm(t)$ and 
$Q(t,r) = \bq(t)$, but equations
(\ref{Mscoeffs}) shows also that 
$\bm$ and $\bq$ are not independent. A~convenient independent
parameter is $\br(t):=R_0>0$, in terms of which we have 
\begin{mathletters}
\label{bmbqbr}
\begin{eqnarray}
\bm
&=&
\br 
\! 
\left( 1 + 2\br^2\ell^{-2} \right)
\ \ ,
\\
\bq
&=&
\epsilon 
\br
\left( 1 + 
3 \br^2 \ell^{-2} \right)^{1/2}
\ \ ,
\end{eqnarray}
\end{mathletters} 
where $\epsilon$ is a discrete parameter taking the
values~$\pm1$. The reduced action reads 
\begin{equation}
S 
=
\int dt
\left(
{\bf p}_\br {\dot \br}
- {\bf h} 
\right)
\ \ ,
\label{S-red}
\end{equation}
where
\begin{equation}
{\bf p}_\br = 
\left( 1 + 
6 \br^2 \ell^{-2} \right)
\left( \int_0^\infty dr \, P_M \right)
+ 
\frac{ \epsilon \left( 1 + 6 \br^2 \ell^{-2} \right)}
{\left( 1 + 3 \br^2 \ell^{-2} \right)^{1/2}}
\left( \int_0^\infty dr \, P_Q \right)
\ \ ,
\end{equation}
and the reduced Hamiltonian ${\bf h}$ is
\begin{equation}
{\bf h} = 
\epsilon 
\left( {\tilde \Phi}_+ - {\tilde \Phi}_0 \right) 
\br \! \left( 1 + 
3 \br^2 \ell^{-2} \right)^{1/2}
+ {\tilde N}_+ 
\br 
\! 
\left( 1 + 
2 \br^2 \ell^{-2} \right)
\ \ .
\label{bfh-def}
\end{equation}
${\bf p}_\br$ can be interpreted geometrically in terms of the Killing
time 
evolution and the electromagnetic gauge choice at the two ends of the
constant $t$ hypersurfaces as in Ref.\ \cite{lou-win}, and 
the dynamics derived from ${\bf h}$ can be verified to have the correct
geometric content.

We note that it would be possible to do the reduction in two stages,
imposing in the first stage all the constraints {\em except\/} 
what setting $M_1$ (\ref{Mscoeff1}) to zero 
implies for the interdependence of 
$\bm$ and~$\bq$. 
After this first stage, one arrives at 
a four-dimensional phase space on which $\bm$ and $\bq$ and 
their conjugate momenta, found as in Ref.\ \cite{lou-win}, provide a
canonical chart. 
The single remaining constraint is an algebraic relation
between $\bm$ and~$\bq$, and it clearly Poisson commutes with the
Hamiltonian, which does not depend on the momenta. 
Elimination of the last constraint then duly leads to
the fully reduced two-dimensional 
phase space found above.\footnote{We thank Bernard
  Whiting for discussions on this point.} 

The reduced Hamiltonian ${\bf h}$ (\ref{bfh-def}) 
depends on ${\tilde N}_+$, 
${\tilde \Phi}_+$, and ${\tilde \Phi}_0$, 
but the action (\ref{S-red})
depends in no way on~${\tilde N}_0$. In particular, the action
(\ref{S-red}) does not contain a horizon term of the kind that
produces the Bekenstein-Hawking entropy upon quantizing the analogous
Hamiltonian 
formulation for nonextremal black holes 
\cite{LW2,lau1,BLPP,lou-win,love-hole,brotz-kiefer,Br,Ki}. 
One is led
to conclude that a Hamiltonian quantization 
of the theory (\ref{S-red}) 
along the lines of the 
nonextremal Hamiltonian quantization in Refs.\ 
\cite{LW2,lau1,BLPP,lou-win,love-hole} 
would lead to a vanishing extremal black hole entropy.

\section{Wave packets in Hamiltonian quantization}
\label{sec:packets}

We now wish to include extremal black holes as a limiting case in a
quantum theory that is initially formulated for nonextremal black
holes, and examine the classical limit of the theory, both for
extremal and nonextremal holes, in terms of wave packets. For
concreteness, in this section we set the cosmological constant to
zero. 

In the Hamiltonian theory for nonextremal holes, we take one end of
the spacelike hypersurfaces to be at the asymptotically flat infinity
and the other end at the horizon bifurcation two-sphere, such that the
hypersurfaces are at each end asymptotic to hypersurfaces of constant
Killing time \cite{lou-win,brotz-kiefer,Br}. In the quantum theory, we
then arrive at plane-wave-like wave functions of the form
\begin{equation}
\Psi_{mq}(\alpha,\tau,\lambda)
= \exp \left[ \frac{i}{\hbar}
\left(\frac{A(m,q)\alpha}{8\pi}-m\tau-q\lambda\right)
           \right]
\ \ ,
\label{psimq}
\end{equation}
where the parameters $m$ and $q$ labeling the plane waves have the
interpretation as the mass and charge: they satisfy $m>|q|$, and we
have $A(m,q) = 4 \pi R^2(m,q)$ and $R(m,q) = m + \sqrt{m^2 -
  q^2}$, so that $A$ is the area and $R$ the area-radius of the
horizon. Of the three arguments $(\alpha,\tau,\lambda)$ of the wave
function, $\tau$~has an interpretation as the Killing time at the
infinity, $\lambda$ is related to the electromagnetic gauge choice
at the infinity and at the horizon, and $\alpha$ is the
rapidity parameter of the normal vector 
to the spacelike hypersurfaces at
the bifurcation two-sphere. 
One way to arrive at the wave functions (\ref{psimq}) is
the leading-order semiclassical approximation to the Wheeler-DeWitt
equation in the metric variables
\cite{brotz-kiefer,Br,Ki}.\footnote{In this case, going to higher
  orders in the semiclassical approximation would change the wave
  function in ways that are important when fields with local degrees
  of freedom come into play; how the Hawking radiation can be obtained
  from the wave functional at the next order was shown in Ref.\ 
  \cite{DK}.}  
Another way is to introduce $\tau$ and $\alpha$ as
reparametrization clocks and perform an exact quantization of the 
Hamiltonian theory along the lines of
equations (191)--(192) of Ref.\ \cite{kuchar1}. 
Note that all the three arguments $(\alpha,\tau,\lambda)$ stand on
an equal footing as ``configuration'' variables, and the wave function 
does not depend on an additional, external ``time'' variable. 

It is useful to point out the parallels between the plane waves
(\ref{psimq}) and the plane wave states for the free nonrelativistic
particle, proportional to $\exp(ikx-i\omega t)$. As $A=A(m,q)$, the
number of parameters in the 
states $\Psi_{mq}$ is one less than the number of arguments; 
the same holds 
for the particle, as there $\omega=\omega(k)=k^2/2m$. The phase of
(\ref{psimq}) is $\hbar^{-1}$ times a particular solution to the
Hamilton-Jacobi equation, labeled by $m$ and~$q$, 
and varying the phase with respect to the
parameters yields the equations \cite{Br,Ki}
\begin{mathletters}
\label{alpha-lambda}
\begin{eqnarray}
\alpha &=& 8\pi\left(\frac{\partial A}{\partial m}\right)^{-1}\tau
        \equiv \kappa\tau \ , 
\label{alpha}
\\
\lambda &=& \frac{\kappa}{8\pi}\frac{\partial A}{\partial q}\ \tau
        \equiv \phi\tau \ , 
\label{lambda}
\end{eqnarray}
\end{mathletters}
where $\kappa$ denotes the surface gravity and $\phi$ the
electrostatic potential difference between the infinity 
and the horizon: these are the equations for 
the family of classical
spacetimes recovered from the particular solution to the
Hamilton-Jacobi equation. 
In comparison, for the free particle the corresponding
extremization yields the particular classical 
trajectory $x=kt/m$. 

Now, if we were to perform a similar quantization for the extremal
holes on their own, the first term in the exponent in (\ref{psimq})
would not be present, as the analysis in section \ref{sec:fall-ex-hor}
shows.  (Note that this is consistent with
equation~(\ref{alpha}), as the surface gravity for the extremal hole
vanishes.)  In the analogy with the free particle, this is as if a
particular value for the momentum, say~$p_0$, were special in the
sense that no dynamical variables $(x,p)$ existed for $p=p_0$.
However, a classical correspondence for the free particle is not
gained from the plane wave solutions itself, but from wave {\em
  packets\/} that are obtained by superposing different wave
numbers~$k$. Only such superpositions yield quantum states that are
sufficiently concentrated near individual classical trajectories, such
as $x=k_0t/m$. We shall proceed similarly with the quantum state
(\ref{psimq}) and first build for nonextremal holes 
wave packets that are concentrated along the
classical relations~(\ref{alpha-lambda}). 
After having built these packets, we then extend them, by hand, to the
extremal limit, and let this limit {\em define\/} 
what we mean by extremal
holes in the quantum theory. 
This procedure might be called
``extremization after quantization", and it mirrors the spirit of the
path integral approach in Ref.\ \cite{GM}.

For the explicit construction of the wave packets, we integrate
over $A$ and $q$ and express the mass $m$ as a function of
these variables,
\begin{equation}
m(A,q)= 
\frac{R}{2} 
\left( 
1 + \frac{q^2}{R^2}
\right)
= 
\frac{A+4\pi q^2}{4\sqrt{\pi A}}
\ \ . 
\label{mAq}
\end{equation}
The integration range is $A>4\pi q^2$. 
For the weight functions we choose Gaussians that are peaked around
the values $A=A_0$ and $q=q_0$: 
\begin{eqnarray}
\psi(\alpha,\tau,\lambda) 
&=& 
\int_{A>4\pi q^2}
dAdq\ 
   \exp\left(-\frac{(A-A_0)^2}{2(\Delta A)^2}-\frac{(q-q_0)^2}
   {2(\Delta q)^2}\right)\nonumber\\
 & & 
\ \ \ 
\times \exp \left[ \frac{i}{\hbar}
\left(\frac{A\alpha}{8\pi}-m(A,q)\tau-q\lambda\right)
           \right]
\ \ .
\label{super}
\end{eqnarray}
Provided $A_0$ and $q_0$ are not close to the extremal limit,
$A_0=4\pi q_0^2$, and provided $\Delta A$ and $\Delta q$ are chosen
suitably, it is a good approximation to expand 
$m(A,q)$ around $A_0$ and $q_0$ to quadratic order and then take the
integral over all real $A$ and~$q$. We denote  
the values of $m$, $\phi$, and $\kappa$ at $(A_0,q_0)$
by $m_0$, $\phi_0$, and~$\kappa_0$, respectively.
The corresponding horizon radius is called~$R_0$.
The calculation is lengthy but straightforward. 
Apart from overall normalization and phase factors, the result reads 
\begin{eqnarray}
\psi(\alpha,\tau,\lambda) 
&=& 
{\cal N} 
\exp \left[ \frac{i}{\hbar}
\left(\frac{A_0\alpha}{8\pi}-m_0\tau-q_0\lambda\right)
           \right]
\nonumber\\
 & & \ \times \exp\left(-\frac{(\lambda-\phi_0\tau)^2}{2\hbar^2}
      \frac{{\cal F}}{{\cal B}}-\frac{(\alpha-\kappa_0\tau)^2}{2\hbar^2}
      \frac{{\cal G}}{{\cal B}}\right)\nonumber\\
 & & \; \times\exp\left(\frac{(\lambda-\phi_0\tau)(\alpha-\kappa_0\tau)}
        {2\hbar^2}\frac{{\cal H}}{{\cal B}}\right)\times
       (\mbox{phase factors})\ , \label{packet}
\end{eqnarray}
where
\begin{mathletters}
\label{BFGH}
\begin{eqnarray}
{\cal B} &=& \left(1+\frac{\tau^2\kappa_0(\Delta A)^2(\Delta q)^2}
  {64\hbar^2\pi^2R_0^3}\right)^2 +\frac{4\pi\tau^2}{\hbar^2A_0}
   \left((\Delta q)^2+ \frac{(\Delta A)^2(1-3\kappa_0R_0)}
   {16\pi A_0}\right)^2\ , \label{B} \\
{\cal F} &=& (\Delta q)^2+ \frac{\tau^2(\Delta A)^2(\Delta q)^2}
  {8\hbar^2A_0^2}\left((\Delta q)^2(1-2\kappa_0R_0)
  +\frac{(\Delta A)^2(1-3\kappa_0R_0)}{8\pi A_0}\right)\ , \label{F} \\
{\cal G} &=& \frac{1}{64\pi^2}\left((\Delta A)^2
  +\frac{4\pi\tau^2(\Delta A)^2(\Delta q)^2}{\hbar^2A_0}
  \left[(\Delta q)^2+\frac{(\Delta A)^2(1-2\kappa_0R_0)}{16\pi A_0}\right]
   \right)\ , \label{G}\\
{\cal H} &=& \frac{\tau^2(\Delta A)^2(\Delta q)^2q_0}{2\hbar^2A_0^2}
  \left((\Delta q)^2+ \frac{(\Delta A)^2(1-3\kappa_0R_0)}{16\pi A_0}
  \right)\ . \label{H}
\end{eqnarray}
\end{mathletters}
The packet is, as expected, concentrated around the classical
values (\ref{alpha-lambda}), but it has -- analogously to
the free particle -- a width that ``spreads" with increasing
time~$\tau$. We note that the term $1-2\kappa_0R_0$ occurring in these
expressions becomes zero for vanishing charge (Schwarzschild case),
while $1-3\kappa_0R_0$ becomes zero for $q_0^2=3m_0^2/4$, which is
the thermodynamical stability boundary for charged black holes 
with fixed charge \cite{davies1}.

Let us pause to comment on the special case of Schwarzschild black
hole. The charge terms are absent, and one is left
with the wave packet
\begin{equation}
\psi(\alpha,\tau)={\cal N} \exp\left(\frac{iA_0\alpha}{8\pi\hbar}
  -\frac{im_0\tau}{\hbar}-\frac{(\alpha-\kappa_0\tau)^2}{2\hbar^2}
   \frac{{\cal G}'}{{\cal B}'}\right)
   \times\mbox{(phase factors)}\ , \label{Sch}
\end{equation}
where
\begin{equation}
{\cal B}'=1+\frac{\tau^2(\Delta A)^4}{256\pi\hbar^2A_0^3}, \quad
{\cal G}'= \frac{(\Delta A)^2}{64\pi^2}\ . \label{prime}
\end{equation}
{}From the expression for ${\cal B}'$ one can read off the time scale,
$\tau_*$, of the spreading:
\begin{equation}
\tau_*= \frac{16\hbar\sqrt{\pi}A_0^{3/2}}{(\Delta A)^2}\ . 
\label{taustar}
\end{equation}
The minimal value for $\Delta A$ should be of the order of the
Planck length squared, i.e., $\Delta A\propto\hbar
\approx 2.6\times 10^{-66}\mbox{cm}^2$. This corresponds to a 
black hole as classical as possible. The corresponding dispersion time
from (\ref{taustar}) is
\begin{equation}
\tau_*=\frac{128\pi^2R_0^3}{\hbar}\approx 10^{73}
         \left(\frac{m_0}{m_{\odot}}\right)^3\mbox{sec}\ . 
\label{evap}
\end{equation}
Note that this is just of the order of the black hole evaporation time!
The occurrence of this timescale is not very surprising, however,
since the evaporation time gives also the timescale for the
breakdown of the semiclassical approximation.

In the general case of nonvanishing charge, the dispersion time
also depends on the charge uncertainty~$\Delta q$. A~direct comparison
with (\ref{evap}) can be made if only the last term in~(\ref{B}) -- the
term that only depends on~$\Delta A$ -- is taken into account: For small
charge, the dispersion time (\ref{evap}) increases according to
\begin{equation}
\tau_*\to\tau_*\left(1+\frac{3q_0^2}{R_0^2}\right) 
\ . 
\end{equation}
This may be interpreted as being due to the fact that the Hawking
temperature for charged holes is smaller than for uncharged ones.
Taking into account also the $\Delta q$-terms in~(\ref{B}), 
the dispersion time generally decreases.
 
Consider now the extremal limit, 
in which the center of the packet is driven to 
$A_0=4\pi q_0^2$. As this center is now on the boundary
of the integration domain in~(\ref{super}), the approximations made
above in the evaluation of the integrals are 
no longer fully justified. 
However, as the integrand in
(\ref{super}) is a smooth function of $A$ and $q$ at and beyond
the boundary, the expressions (\ref{packet}) 
and (\ref{BFGH}) should still remain
qualitatively correct. Assuming this is the case, and recognizing that
$A_0=4\pi q_0^2$ implies $\kappa_0=0$, we see that the widths of the
Gaussians in (\ref{packet}) are $\tau$-{\em independent} for large
enough~$\tau$. 
This is, again, not surprising, since the extremal hole has 
vanishing temperature and thus does not evaporate. For example,
taking the large $\tau$ limit and choosing the minimal widths
$\Delta A\propto\hbar$ and $\Delta q\propto\sqrt{\hbar}$, one finds
that the $\alpha$-dependence of the wave packet is the Gaussian factor
\begin{equation}
\exp\left(-\frac{\alpha^2}{128\pi^2}\right)\ .
\end{equation}
This factor is independent of both $\tau$ and~$\hbar$.

It is evident that although our wave packet for $A_0=4\pi q_0^2$ is
peaked at vanishing~$\alpha$, the packet has support also at
$\alpha\ne0$, and the packet does not seem to be qualitatively 
different from one for which $A_0$ is close to but not
exactly equal to $4\pi q_0^2$. In this approach, one would thus expect
the extremal hole to have the usual Bekenstein-Hawking entropy.

\section{Concluding remarks}
\label{sec:conclusion}

We have discussed two complementary approaches to a Lorentzian
Hamiltonian quantum theory that would encompass extremal black holes.
In our classical Hamiltonian theory comprising only extremal black
holes, the Hamiltonian does not contain a horizon surface term, and it
is difficult to see how quantization of such a theory could lead to a
nonvanishing result for the black hole entropy.
If, on the other
hand, the extremal case is understood as a certain limit in a quantum
theory that encompasses both the extremal and nonextremal cases, such
a term will emerge, since the extremal case is only ``of
measure zero". This becomes especially transparent from the analogy
with the free particle, for which it would seem peculiar to separately
quantize some lower-dimensional set of classical solutions. 
We emphasize again that
the very notion of a black hole is a classical one, analogous to the
notion of a classical trajectory in particle mechanics. 

Although we discussed how the Bekenstein-Hawking 
entropy for an extremal black hole would in principle 
arise from the quantum theory of section~\ref{sec:packets}, 
we did not attempt a direct computation of the entropy. 
One might expect that the entropy could be recovered as an
``entanglement entropy'' between the wave packets built 
in section~\ref{sec:packets}. 
Examining this question lies, however, 
beyond the scope of this paper. 

We have throughout the paper understood the extremal limit of
nonextremal black holes so that the structure of the asymptotic
infinity remains qualitatively unchanged. This meant that the limiting
spacetime is indeed a black hole, and the horizon structure
experiences a qualitative change in the limit. However, it is possible
to take the extremal limit also in a way that preserves a bifurcate
Killing horizon, at the cost of having the infinity undergo a
qualitative change. The resulting spacetimes are of the
Bertotti-Robinson type, in which the radius of the two-sphere
is constant on the spacetime
\cite{exact-book,zaslavskii-extr1,zaslavskii-extr2,bri-lo-pe}. 
These spacetimes are geodesically complete and cannot be interpreted
as black holes, and the horizon is an acceleration horizon rather than
an event horizon. Nevertheless, if one works under boundary conditions
that do not require an infinity, 
for example by taking the ``outer''
end of the spacelike hypersurfaces to be at a finite ``box'', 
it is possible to include the Bertotti-Robinson type 
spacetimes in a Hamiltonian formalism
that handles the horizon bifurcation two-sphere as in Refs.\ 
\cite{LW2,lau1,BLPP,lou-win,love-hole,brotz-kiefer,Br,Ki}. As a
result, one finds that the acceleration horizon is associated with an
entropy equal to one quarter of the area
\cite{zaslavskii-extr1,zaslavskii-extr2}. This is similar to the
result for the acceleration horizon in Rindler spacetime 
\cite{laf-flat-entropy}. 

We wish to add here some remarks on the situation in
string theory (for a recent review, see \cite{Peet}).
The issue of black hole entropy is also there addressed in
the framework of a semiclassical approximation: while the
semiclassical approximation used in section \ref{sec:packets} 
can be found through an expansion
with respect to the gravitational constant \cite{Ki}, 
the semiclassical approximation in string theory is accomplished 
through an expansion with respect to the string length.
In addition, one can vary the string coupling constant at the
level of the effective action and thereby connect the large-coupling
regime of black holes with the small-coupling regime of D-branes
in Minkowski spacetime. From a methodological point of view,
string theory employs the
``quantization before extremization"-method used in the wave packet
construction of section~\ref{sec:packets}.
For the purpose of calculating the entropy, one looks in the
quantum theory for states that are eigenstates of the Hamiltonian
and some gauge generator with respective eigenvalues $m$
(mass) and $q$ (generalized charges); the so-called BPS states
are then defined as the states in the ``small representation" for which
$m=\vert q\vert$, giving the condition of extremality.
In principle one should 
be able to perform superpositions and construct wave packets in the
manner of section \ref{sec:packets} also in string theory, 
although this has, to our knowledge, 
not been done.

An interesting open problem is the possible occurrence of a naked
singularity. The boundary conditions of 
section \ref{sec:fall-ex-hor} clearly 
do not comprise a singular three-geometry. 
In the formulation of section~\ref{sec:packets}, however, 
the wave packet (\ref{packet}) contains also parameter values that
would correspond to such singular geometries. 
These geometries could be
avoided if one imposed the boundary condition that the wave function
in the momentum representation vanish for such values. 
Continuity would then also enforce that the wave function vanish
at the boundary itself, i.e., for the extremal case. Consequently,
quantum gravity would forbid the existence of extremal holes!
Such a consequence would also follow in string theory. 
This is certainly an interesting aspect that should deserve
further investigation.

\acknowledgments
We thank Don Marolf 
for raising the issue of internal infinity boundary conditions in
the Hamiltonian evolution of an extremal black hole, and for
discussions. 
We also thank Bernard Whiting for his comments on 
an early version of the manuscript. 
C.~K. would like to
thank the Max-Planck-Institut f\"ur Gravitationsphysik 
for hospitality at the early stage of this work.

\appendix
\section*{Extremal
Reissner-Nordstr\"om-anti-de~Sitter black holes}
\label{app:rnads}

In this appendix we recall some relevant properties of the extremal
\rn\ and \rnads\ metrics. 

In the curvature coordinates $(T,R)$, the \rnads\ (RNAdS) metric reads 
\begin{mathletters}
\label{rnads-metric}
\begin{equation}
ds^2 = - F dT^2 + F^{-1} dR^2 + R^2 d\Omega^2
\ \ ,
\label{curv-metric}
\end{equation}
where $d\Omega^2$ is the metric on the unit two-sphere and
\begin{equation}
F := {R^2 \over \ell^2} + 1 - {2M \over R} + {Q^2 \over R^2}
\ \ .
\label{rnads-F}
\end{equation}
\end{mathletters}
We take  $M$ and $Q$ to be real and $\ell>0$. Together with the
electromagnetic potential
\begin{equation}
A = {Q \over R} dT
\ \ ,
\label{rnads-A}
\end{equation}
the metric~(\ref{rnads-metric}) is a solution to the Einstein-Maxwell
equations with the cosmological constant
$-3\ell^{-2}$ \cite{exact-book,hoffmann}. The parameters $M$ and $Q$
are referred to respectively as the mass and the (electric) charge.

We are interested in the case where the quartic polynomial $R^2 F(R)$
has a positive double root, $R=R_0$, such that $F$ is positive for
$R>R_0$. The necessary and sufficient condition for this to happen is
that $Q\ne0$ and $M=M_{\rm crit}(Q)$, where
\begin{equation}
M_{\rm crit}(Q) := {\ell \over 3 \sqrt{6} }
\left( \sqrt{1 + 12 {(Q/\ell)}^2} +2 \right)
{\left( \sqrt{1 + 12 {(Q/\ell)}^2} -1 \right)}^{1/2}
\ \ .
\label{McritQ}
\end{equation}
We then have $R_0 = \Rcrit(Q)$, where 
\begin{equation}
\Rcrit(Q) :=
{\ell\over\sqrt{6}}
\left( \sqrt{1 + 12 {(Q/\ell)}^2} -1 \right)^{1/2}
\ \ .
\label{Rcrit}
\end{equation}
The metric is uniquely determined by the value of~$Q\ne0$, 
or alternatively by 
the value of $R_0>0$ and the sign of~$Q$. The region 
$R_0 < R < \infty$ covers one exterior region of the extremal RNAdS
hole. The Penrose diagram of the maximal analytic extension can
be found in Refs.\ \cite{lake1,btz-cont}. 

The extremal \rn\ metric is obtained from the extremal RNAdS metric in
the limit $\ell\to\infty$, in which case (\ref{McritQ}) and
(\ref{Rcrit}) reduce to $M_{\rm crit}(Q) = \Rcrit(Q) = |Q|$. The
region $R_0 < R < \infty$ covers one exterior region, and the 
Penrose diagram of the maximal analytic extension can
be found for example in Ref.\ \cite{haw-ell}.

\end{document}